\documentclass[useAMS,usenatbib]{mn2e}

\usepackage{natbib}
\usepackage{epsfig}
\usepackage{calc}
\usepackage{amssymb}
\usepackage{amstext}
\usepackage{amsmath}
\usepackage{multicol}
\usepackage{pslatex}
\usepackage{fancyhdr}
\usepackage{longtable}
\usepackage{pdflscape}
\usepackage{graphics}

\title{
{Bright low mass eclipsing binary candidates observed by STEREO}
}

\author[K. T. Wraight, L. Fossati, Glenn J. White, A. J. Norton and D. Bewsher]{K. T. Wraight$^{1}$\thanks{e-mail:
k.t.wraight@open.ac.uk}, L. Fossati$^{1}$, Glenn J. White$^{1,2}$, A. J. Norton$^{1}$ and D. Bewsher$^{3}$\\
$^{1}$Department of Physical Sciences, The Open University, Milton Keynes, MK7 6AA, UK\\
$^{2}$Space Science and Technology Department, STFC Rutherford Appleton Laboratory, Chilton, Didcot, Oxfordshire, OX11 0QX, UK\\
$^{3}$Jeremiah Horrocks Institute, University of Central Lancashire, Preston, Lancashire, PR1 2HE, UK}

\begin{document}

\pagenumbering{arabic}
\maketitle

\begin{abstract}
Observations from the Heliospheric Imagers (HI-1) on both the \textit{STEREO} spacecraft have been analysed to search for bright low mass eclipsing binaries (EBs) and potential brown dwarf transits and to determine the radii of the companions.  A total of 9 EB candidates have been found, ranging in brightness from $V=6.59$~mag to $V=11.3$~mag, where the radius of the companion appears to be less than $0.4R_{\odot}$, with a diverse range of host temperatures, from 4074~K to 6925~K.  Both components of one candidate, BD-07~3648, appear to be less than $0.4R_{\odot}$ and this represents a particularly interesting system for further study.  The shapes of the eclipses in some cases are not clear enough to be certain they are total and the corresponding radii found should therefore be considered as lower limits.  The EBs reported in this paper have either been newly found by the present analysis, or previously reported to be eclipsing by our earlier \textit{STEREO}/HI-1 results.  One of the new objects has subsequently been confirmed using archival SuperWASP data.  This study was made possible by using an improved matched filter extraction algorithm, which is described in this paper.
\end{abstract}

\begin{keywords}
methods: data analysis -- binaries: eclipsing -- space vehicles: \textit{STEREO} -- stars: brown dwarfs
\end{keywords}

\sloppy

\section{\uppercase{Introduction}}
\label{sec:introduction}

\noindent
Observations of bright host stars of low mass eclipsing companions facilitates precise measurements of the mass and radius of the stars in these systems \citep{johnson2011}.  There are, unfortunately, just 18 detached, double-lined eclipsing binary systems known that include a component with less than 0.8~$M_{\odot}$ \citep{feiden2012}.  The primary objective of this work was to search for previously unknown low-mass EBs from \textit{STEREO}/HI-1 data and to determine the radii of the companions.  This could not be attempted in the earlier analysis of \citet{wraight2011} as it was only through the subsequent development of a matched filter algorithm (see Section \ref{sec:mfa}) that the difficulties of using the \textit{STEREO} HI-1B data (see Section \ref{sec:hi1bodd}) could be overcome, which facilitates the extractions of low amplitude signals from the combined data, which together provide improved phase coverage across the variability cycle.  Compared to our earlier analysis reported in \citet{wraight2011}, updated flatfields had since become available \citep{bewsher2010stereo} and an additional year of data from both satellites for each star had been taken, providing a baseline of about four and a half years compared to the two or three years in our previous study.  Since the initial work of \citet{wraight2011} was restricted to analysing each year of data separately using \textit{STEREO} HI-1A data only, improved accuracy of eclipse depths and periods could be obtained.  The analysis with the matched filter algorithm has allowed us to identify additional candidates that had been missed by the previous search.  In total, nine candidates are presented where the companions appear to be less than 0.4~$R_{\odot}$ and where the eclipses are most likely total and with a low risk of contamination from nearby stars, although the potential for the eclipses to be grazing remains in some cases.

Studies of very low mass stars close to and below the boundary where the interior is fully convective are limited by a lack of bright eclipsing systems involving such stars \citep{feiden2012}.  Where these systems are bright enough, spectroscopic follow-up observations can unambiguously reveal the masses and radii of both components, allowing other parameters to be calculated with sufficient precision to challenge models of stellar formation and evolution \citep{shkolnik2008,cakirli2012,luhman2012,triaud2012}.  Indeed, there is an apparent tendency for current models to under-estimate the radii of low-mass stars, which suggests the presence of a mechanism inflating their radii \citep{cakirli2010,morales2010,morales2011}.  It is therefore highly desirable to have a diverse sample of bright, low-mass detached EBs to compare against these models.

The primary mission of NASA's twin \textit{STEREO} spacecraft is to study the Sun's corona in three dimensions \citep{kaiser2008stereo}.  Data from the Heliospheric Imager on the \textit{STEREO}-Ahead spacecraft \citep[\textit{STEREO}/HI-1A; ][]{eyles2009stereo} has previously been used to produce a catalogue of EBs \citep{wraight2011}, to study the rotational periods of magnetic chemically peculiar stars \citep{wraight2012} and the pulsational periods of long period variables \citep{wraight2012b}, which was possible owing to the stability of the photometry and the accuracy of the calibration \citep[Brown, Bewsher \& Eyles 2009, ][Bewsher, Brown \& Eyles 2012]{bewsher2010stereo}.  Several new EBs from \citet{wraight2011} were suspected to be low mass and confirmation of this provided the motivation to search through data from both the Ahead \textit{STEREO} spacecraft (\textit{STEREO} HI-1A) and the Behind spacecraft (\textit{STEREO} HI-1B) using a custom matched filter algorithm to extract weak variable signals.  

This paper describes the matched filter algorithm used to carry out the large-scale analysis of \textit{STEREO}/HI-1 data in Section \ref{sec:methods}.  The candidate low mass EBs are presented in Section \ref{sec:results}.  In Section \ref{sec:discussion} the sample is discussed in the context of current research in this field.

This sample of low mass EB candidates represents an example of one limited study of \textit{STEREO} data.  There are numerous possibilities for this data and the prospects for the legacy of \textit{STEREO} observations are huge, with a large coverage on the sky of bright stars, regardless of spectral type.

\section{\uppercase{Data analysis}}
\label{sec:methods}

\begin{figure}
\resizebox{4cm}{!}{\includegraphics{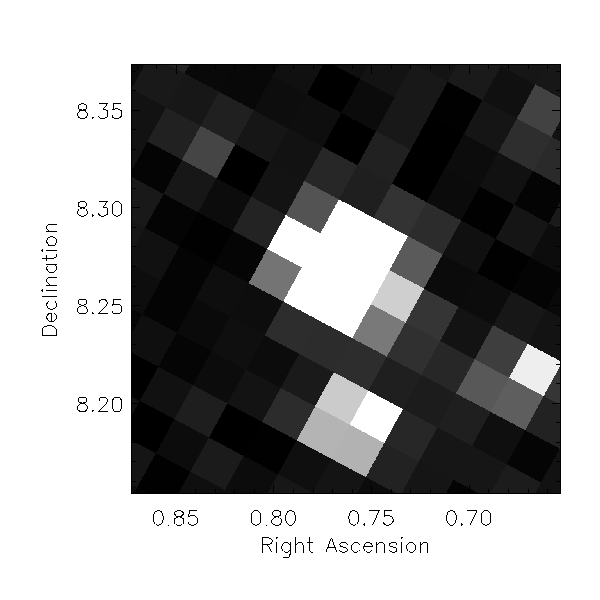}}
\hfill
\resizebox{4cm}{!}{\includegraphics{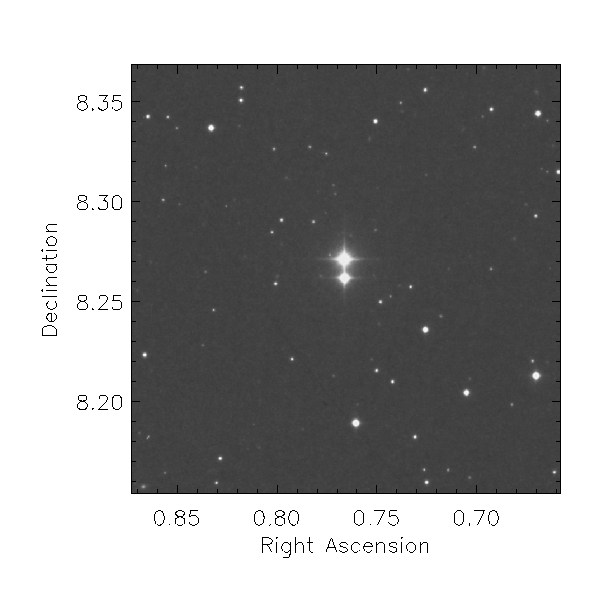}}
\caption{\textit{STEREO}/HI-1A (left) and \textsc{POSSI} (right) views of HIP~247 and HIP~248, illustrating the difficulties of blending with the 70~arcsec pixel scale of \textit{STEREO}/HI-1.}
\label{fig1}
\end{figure}

\noindent
Data from the Heliospheric Imagers (\textit{STEREO}/HI-1) \citep{eyles2009stereo} on both \textit{STEREO} spacecraft was used for the analysis.  A matched filter algorithm (see Section \ref{sec:mfa}) was developed to search through the lightcurves of all of the stars in the \textit{STEREO} database.  The reason a matched filter was used was to allow a good signal to be obtained from a variety of variability types \citep{jenkins1996,weldrake2005,enoch2009,enoch2012}.  However, due to blending with nearby stars, the region close to the Galactic Centre was excluded from the search.  The pixel scale of the \textit{STEREO}/HI-1 imagers is 70 arcsec per pixel \citep{eyles2009stereo} and close to the Galactic Centre, the high density of stars on the sky results in the eclipse depth to be less than the real depth because of blending in the same and/or from neighbouring pixels.  To illustrate this, Figure \ref{fig1} shows HIP~247 and HIP~248, as seen by \textit{STEREO}/HI-1A and compared with an image of the same stars from the \textsc{POSSI} survey.  It is not possible in this particular case to determine which of the two stars is the source of the observed shallow eclipses \citep{wraight2011} from \textit{STEREO} data alone and several other candidates were similarly found to be unresolvable, necessitating follow-up measurements to confirm the companions' status as low mass EBs (most notably BD+03~263p / BD+03~263s and BD+03~2482 / BD+03~2483A).  The matched filter algorithm we have used was developed to search for both general variability and detached and semi-detached EBs in stars brighter than $R=11$th magnitude.

Where available, archival SuperWASP \citep{pollacco2006superwasp} observations of these candidates were checked, to help determine whether any variability was due to a background EB, as additional confirmation of the primary eclipse depth and to clarify whether or not secondary eclipses were present.  Whilst the pixel scale of the Heliospheric Imagers may result in blending in several of the candidates, the eclipse depths and shapes still suggest the presence of a low mass companion.  Additional observations will be required to unambiguously rule out the possibility of grazing eclipses.  Furthermore, radial velocity measurements are required to confirm the low-mass nature of the companions.  We do however note that a significant amount of radial velocity data already exists for HD~75767 \citep{griffin2006} and HD~213597 \citep{nordstroem1997}, both sets of data supporting our suggestion that these are low mass binary systems.

\subsection{Characteristics of \textit{STEREO} data}
\label{sec:hi1bodd}

\noindent
The \textit{STEREO} Heliospheric Imagers \citep{eyles2009stereo} and their calibration \citep{brown2009stereo,bewsher2010stereo,bewsher2012} are summarised here.  The \textit{STEREO}/HI-1 imagers have a field of view of 20 degrees by 20 degrees, which is imaged onto a 2048$\times$2048 element CCD, then binned on-board 2$\times$2 to produce the final 1024$\times$1024 images with 70$\times$70 arcsec pixels.  The spectral response of the filter is peaked between 630~nm and 730~nm (closely resembling a conventional \textit{R}-band filter), although it does allow some light through close to 400~nm and 950~nm \citep{eyles2009stereo}.  Stars of 3rd magnitude or brighter were excluded from the survey, whilst stars down to $R=12$th magnitude from the NOMAD catalogue \citep{zacharias2004} were used as the basis for selection.  Over the course of an orbit, almost 900,000 stars of 12th magnitude and brighter are imaged within 10 degrees of the ecliptic plane.  Since the HI-1 imager field of view is centred only 14 degrees away from the centre of the Sun, data on the sunward side of the field of view occasionally can be affected by solar activity and this region was masked out from the analysis.

The \textit{STEREO}-Ahead spacecraft is in an Earth-leading orbit with a semi-major axis of 0.95 AU, whilst the \textit{STEREO}-Behind spacecraft is in an Earth trailing orbit with a semi-major axis of 1.05 AU \citep{kaiser2008stereo}.  This results in stars remaining in the field of view of the \textit{STEREO}/HI-1A imager for just over 19~days and in the field of view of the \textit{STEREO}/HI-1B imager for just over 22 days, with new observations being recorded every 40~minutes.  With this combination of regular photometry repeated every orbit, the \textit{STEREO} mission is therefore very well suited to a wide variety of stellar variability studies requiring continuous observations over many weeks \citep{wraight2011}.  For this analysis, four such passes of a region of the sky from each satellite were available for most stars, resulting in many thousands of data points spanning up to nearly four and a half years.  With more than twice the amount of data and nearly double the time-span used for the analysis of \citet{wraight2011}, it was possible to extract a larger sample of low mass EB candidates.  The tolerance of the matched filter algorithm to artefacts in the data also allowed the complete set of data for each star to be analysed together, whereas those used in \citet{wraight2011} were restricted to separately using data from each year's observations in isolation to each other.  Improved flatfields also became available \citep{bewsher2010stereo}, improving the efficacy of the data and minimising the amount of detrending of the data that was necessary.

\begin{figure*}
\centering
\resizebox{16cm}{!}{\includegraphics{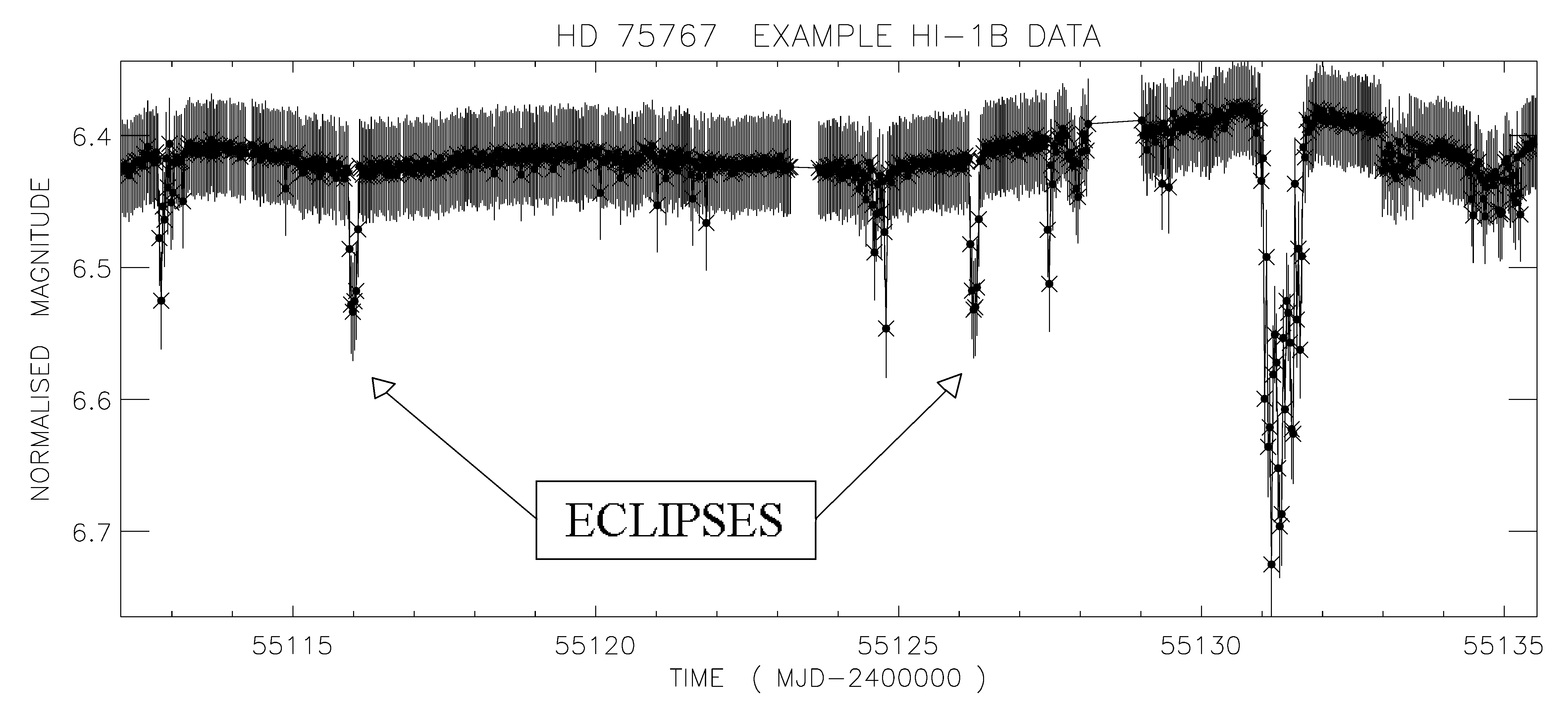}}
\caption{One block of data from \textit{STEREO}/HI-1B of the bright low-mass EB HD~75767.  The eclipses are labelled but all the other sharp decreases in brightness are believed to be due to de-pointing events resulting from micrometeorite impacts.  The smoother changes in brightness are a systematic effect caused by the polynomial detrending.  The error bars shown are the 1~$\sigma$ errors on the individual data points.  A block of data such as this would typically be completely excluded when doing an individual analysis, however as the eclipses are clearly evident a limited section can be considered for inclusion.}
\label{fig2}
\end{figure*}

The \textit{STEREO}/HI-1B data often exhibits sudden decreases in amplitude of all signals at a given time (Figure \ref{fig2}).  These are attributed to errors in the background subtraction due to pointing changes resulting from micrometeorite impacts, as the \textit{STEREO}/HI-1B imager is facing the direction of travel along the spacecraft's orbit, recently confirmed by \citet{davis2012}.  The severity of events can range from a single observation point to days of erratic behaviour.  Often there will be a marked decrease for some hours and then a return to normal.  In spite of the potential for misclassification of such signals as eclipses, they can normally be easily identified by visual examination, as the shape and duration of eclipses in a real EB do not change, beyond the phase coverage and accuracy of the observations, whereas these artefacts are erratically shaped and, importantly, aperiodic.  However, such effects are relatively uncommon in \textit{STEREO}/HI-1A data, and also less severe when they are present, making these observations much more stable than those of \textit{STEREO}/HI-1B.  The matched filter algorithm developed for the analysis is not readily confused by aperiodic changes, since the signal strength found does not usually change for different periods.  However, occasional extreme events cause problems for the polynomial detrending used to clean the lightcurve before being passed to this algorithm (see Section \ref{sec:mfa}, below).

\subsection{Methods used in the analysis}
\label{sec:mfa}

\noindent
Before being passed to a custom matched filter algorithm for analysis, all lightcurves were passed through a culling routine to remove some of the artefacts caused by Mercury and Venus passing through the field of view as well as some of the more extreme pointing-related events, although as it is intended to leave some variability untouched, such as from eclipses, this is not wholly effective.  Points more than 4$\sigma$ away from the weighted mean magnitude, for each epoch, were excluded in this process.  Polynomial detrending was then carried out by a 4th order polynomial in order to remove residual trends from the flat fielding or other artefacts.  This also removes some variability that is long compared to the length of an epoch, thus periods longer than about half the length of an epoch are not expected to be reliable, however long period eclipses are not removed in this process.

\begin{figure*}
\centering
\resizebox{5cm}{!}{\includegraphics{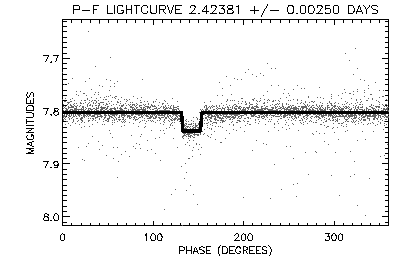}}\hfill
\resizebox{5cm}{!}{\includegraphics{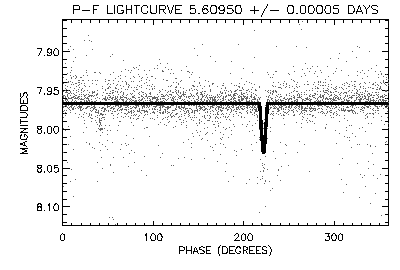}}\hfill
\resizebox{5cm}{!}{\includegraphics{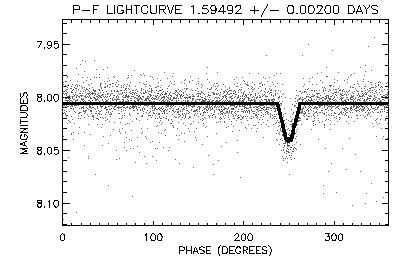}}
\caption{The different eclipse shapes utilised by the matched filter algorithm described in Section \ref{sec:mfa}.  The EBs are HD~213597 (box type, left), V818~Tau (v-shaped, middle, note that secondary eclipses are just visible) and HD~222891 (composite  with one-tenth of the eclipse in totality, right).}
\label{fig3}
\end{figure*}

The matched filter algorithm analyses a lightcurve in several stages, outlined below, and works by building model lightcurves and finding the minimum least-squared error of a model compared to the actual lightcurve.  The model lightcurves consist of data points with the same time and errors as the real lightcurve.  The sequence of the processing steps is as follows:

\begin{enumerate}
\item Determine a best-fitting period.  A sinusoidal shape is used at this stage.  A periodogram is produced during this process.
\item Fine-tune the period.  A precise period is the most important step in the process and without this the other characteristics will not be reliably determined.
\item Determine the amplitude of variability.  This needs to be done before and after determining the best-fitting shape.
\item Determine the shape of the variability.  A selection of shapes consisting of sinusoidal variability with different harmonic signals overlaid are used in addition to shapes based upon box-like total eclipses, v-shaped eclipses and a composite of total eclipses with wide ingress and egress phases.  The amplitude is recalculated after this stage and, if an eclipsing model was best-fitting, the duration and depth of eclipses.  Figure \ref{fig3} shows the different eclipsing models as applied to EBs either from the present sample or from previously known sources (it does not produce a v-shaped model for any EB in the sample).
\item If the best-fitting amplitude is zero this means that the model finds the star to be constant.  The period will still be returned.
\item If an eclipsing model was best-fitting and the amplitude non-zero, eccentricity and amplitude of secondary eclipses is checked for.  Higher harmonics of the period are also checked.
\end{enumerate}

This matched filter algorithm is processor-intensive and, depending on the number of data points, may take from 5 to 10 minutes per star to process for a period range from $0.1$ to $3.5$ days.  This has a resolution of $0.005$ days initially and then is fine-tuned to $0.00005$ days in a narrow search around the best-fitting period.  Higher frequencies were excluded to avoid the Nyquist frequency of about $15.625~\mbox{days}^{-1}$ ($0.064~\mbox{days}$).  This was the period range used in the search, however the program showed during testing a clear preference for finding the half-period harmonic of EBs, to the extent that it was specifically programmed to check at double and triple the periods initially found to see if these produced a better fit.  It thus potentially checked periods up to 10.5~days for eclipses.

A table of statistics was produced after analysing each field of view.  These tables could then be statistically analysed to try to identify low mass EB candidates.  Visual examination of the periodograms produced by the matched filter, along with lightcurves phase-folded on the best-fitting period and the detrended lightcurves themselves, was then undertaken to produce a shortlist of signals for further examination.

After the initial list of candidates had been compiled, the surrounding stars in the vicinity of low mass EB candidates were checked to assess whether there was likely to be any contamination from other nearby known bright EBs.  This ruled out many candidates.  Those remaining were cross-checked in the literature to see if they were already known prior to any analysis with \textit{STEREO}.  Archival SuperWASP \citep{pollacco2006superwasp} data was available for some candidates and this was further checked to confirm both the source and depth of eclipses.  A more detailed study of the known parameters for each star was carried out using information available in the literature, which showed one candidate to be likely to be a giant and the companion therefore too large to be a low mass EB.  A handful of the remaining candidates were then set aside for potential follow-up but excluded from presentation here, as the source of eclipses needed to be more reliably determined or the eclipses appeared more likely to be grazing.  These candidates were subjected to the same analysis as the nine that were left, however the analysis is of more limited use when the shape and depth of eclipses is not reliably determined (in particular, two apparently promising new candidate EBs, HD~18731 and HD~73470, appeared more likely to be grazing).

\section{\uppercase{Low Mass EB candidates}}
\label{sec:results}

\subsection{Lightcurves and data}

\begin{figure*}
\centering
\resizebox{16cm}{!}{\includegraphics{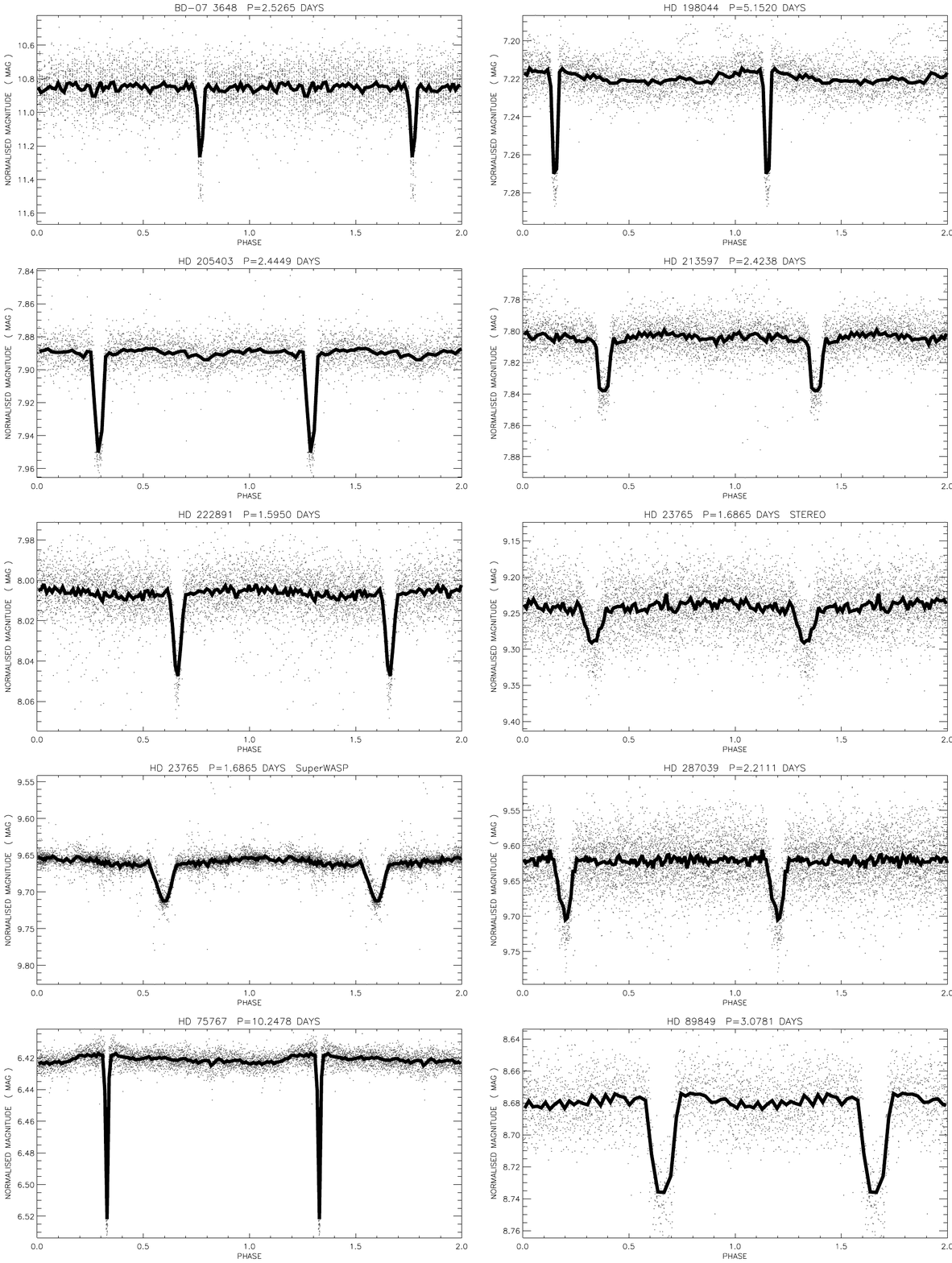}}
\caption{Phase-folded lightcurves of the low mass EB candidates found from the analysis of \textit{STEREO}/HI-1 data (see Section~\ref{sec:results}) and SuperWASP lightcurve (4th row, left column) of one of the candidates phase-folded on the same period as the \textit{STEREO}/HI-1 counterpart (HD~23765, 3rd row, right column).  The best-fitting lines shown bin 40 data points together with outliers more than 3~$\sigma$ from the mean of each bin culled.  Two cycles are shown.}
\label{fig4}
\end{figure*}

\begin{figure*}
\centering
\resizebox{3cm}{!}{\includegraphics{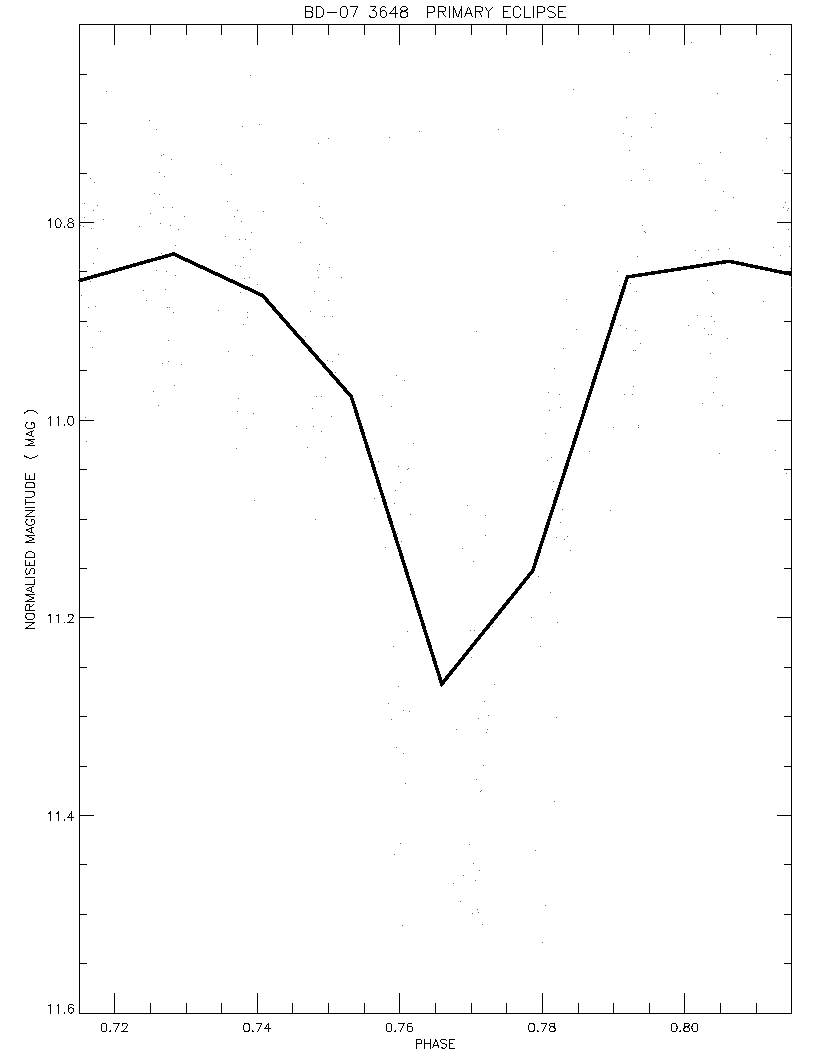}}
\resizebox{3cm}{!}{\includegraphics{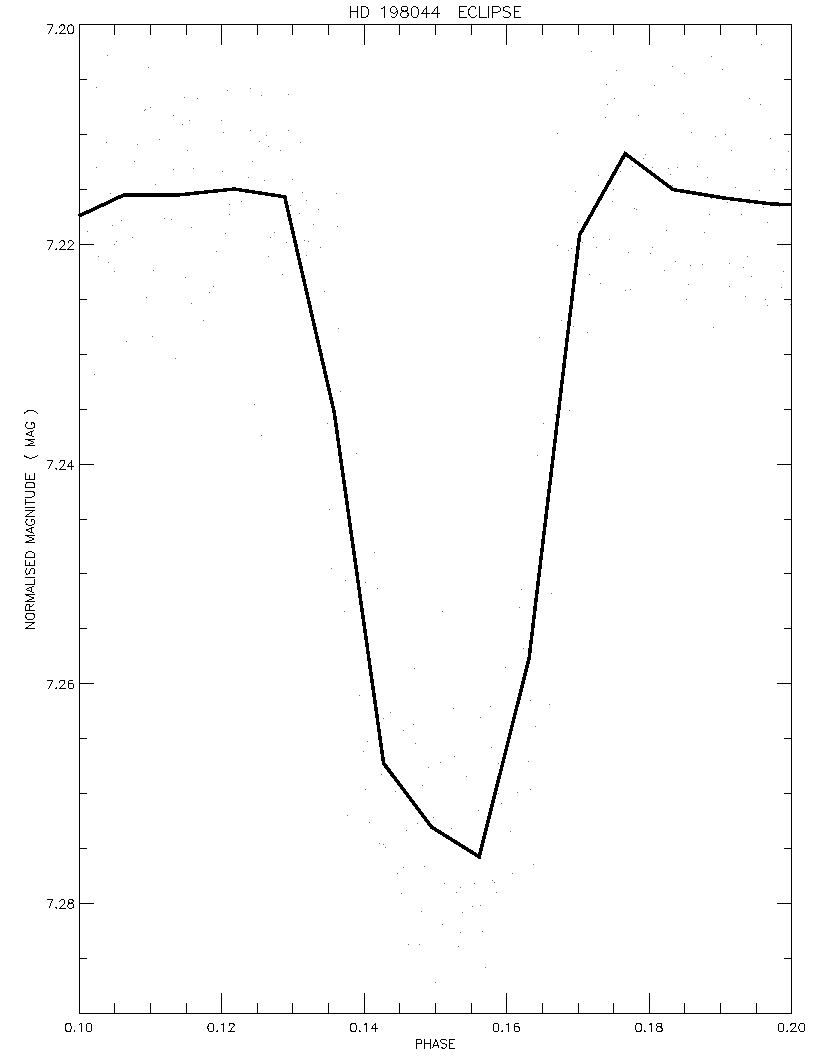}}
\resizebox{3cm}{!}{\includegraphics{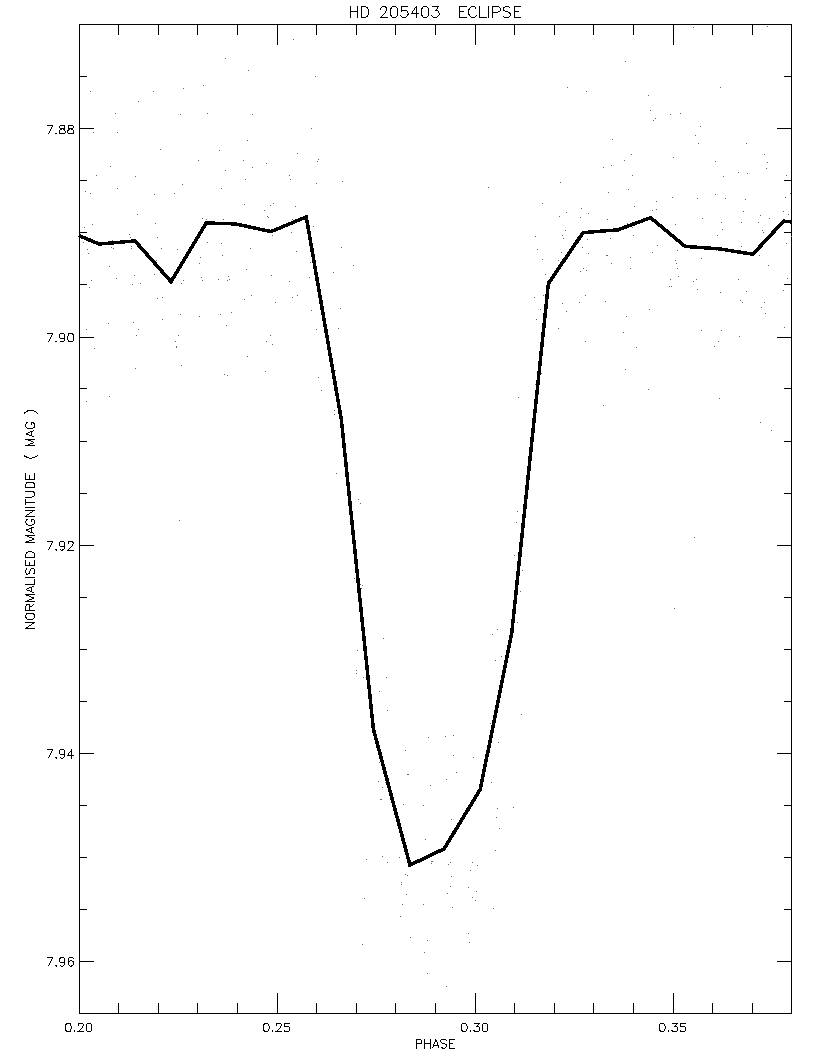}}
\resizebox{3cm}{!}{\includegraphics{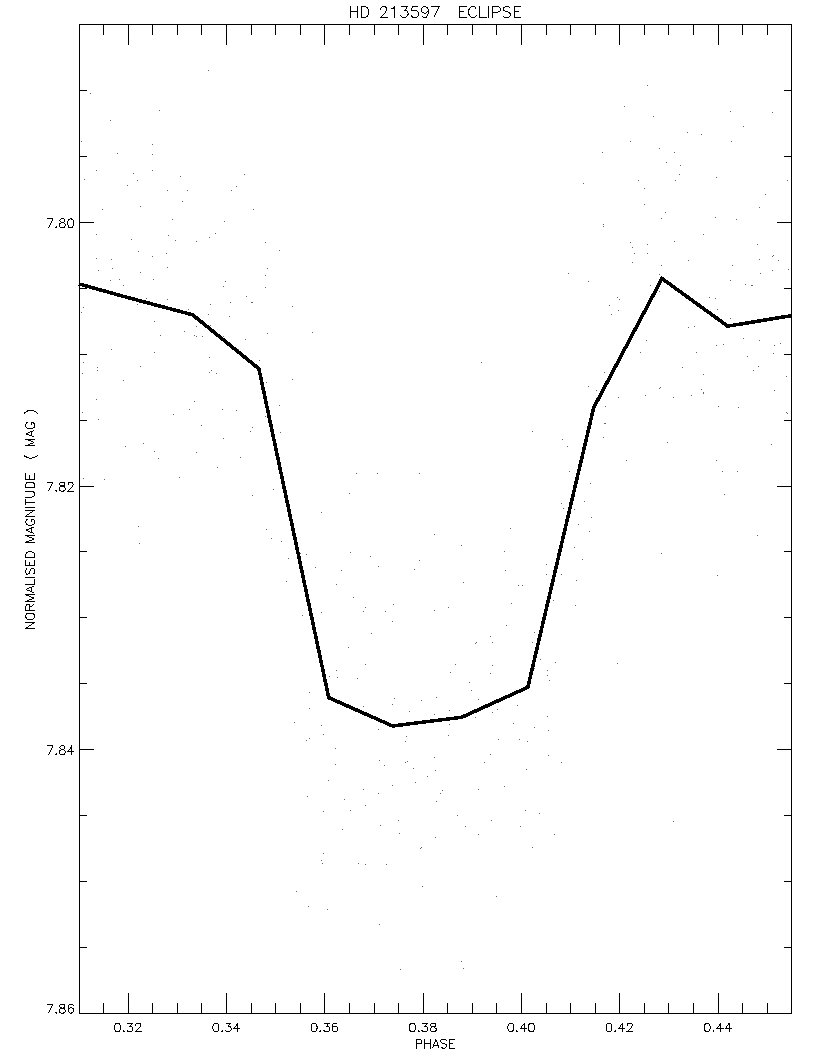}}
\resizebox{3cm}{!}{\includegraphics{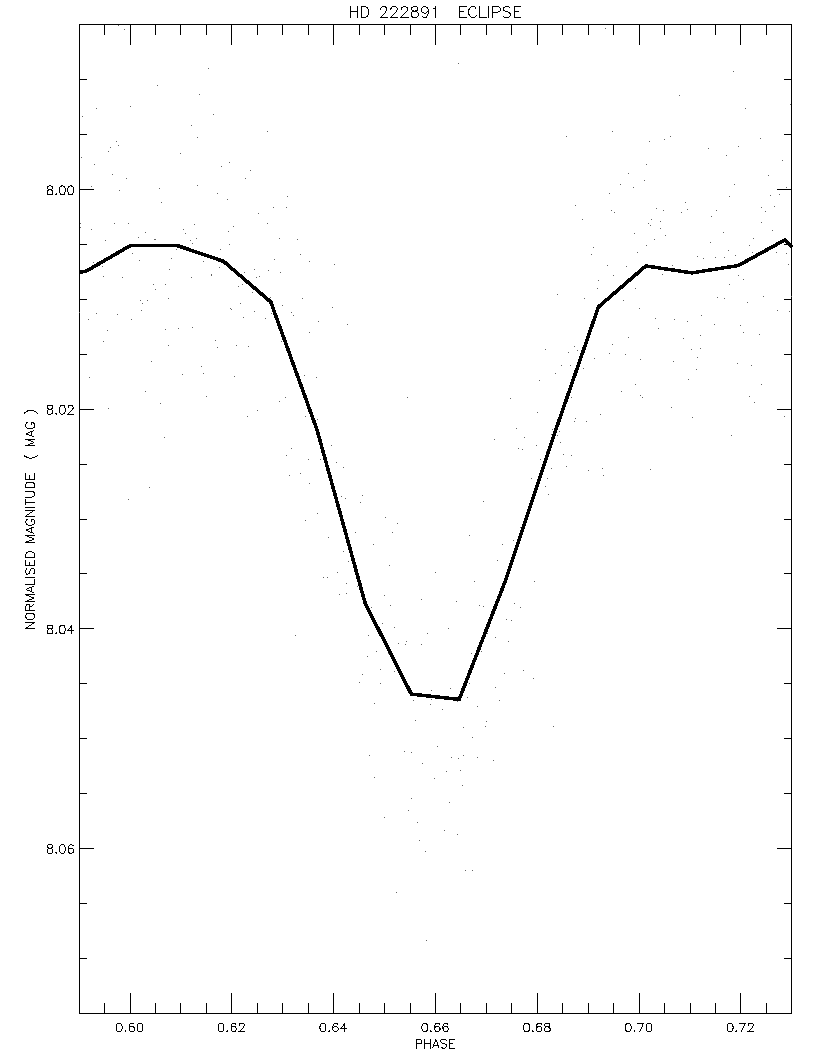}}\\
\resizebox{3cm}{!}{\includegraphics{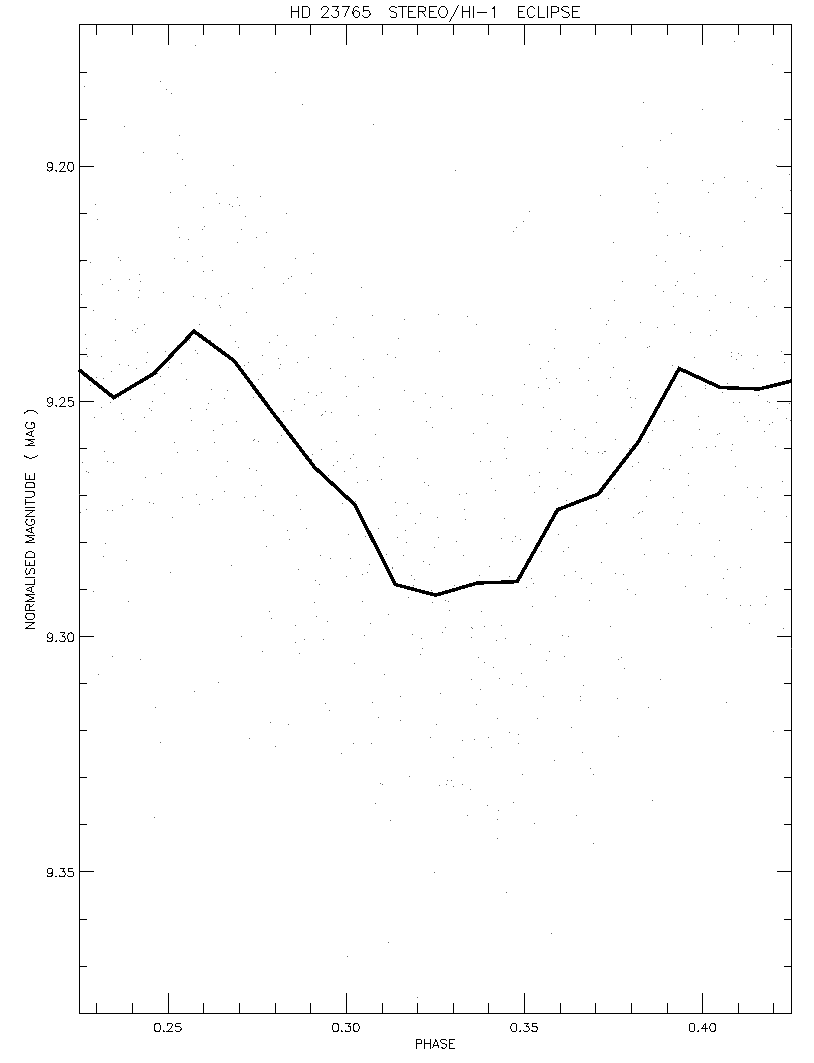}}
\resizebox{3cm}{!}{\includegraphics{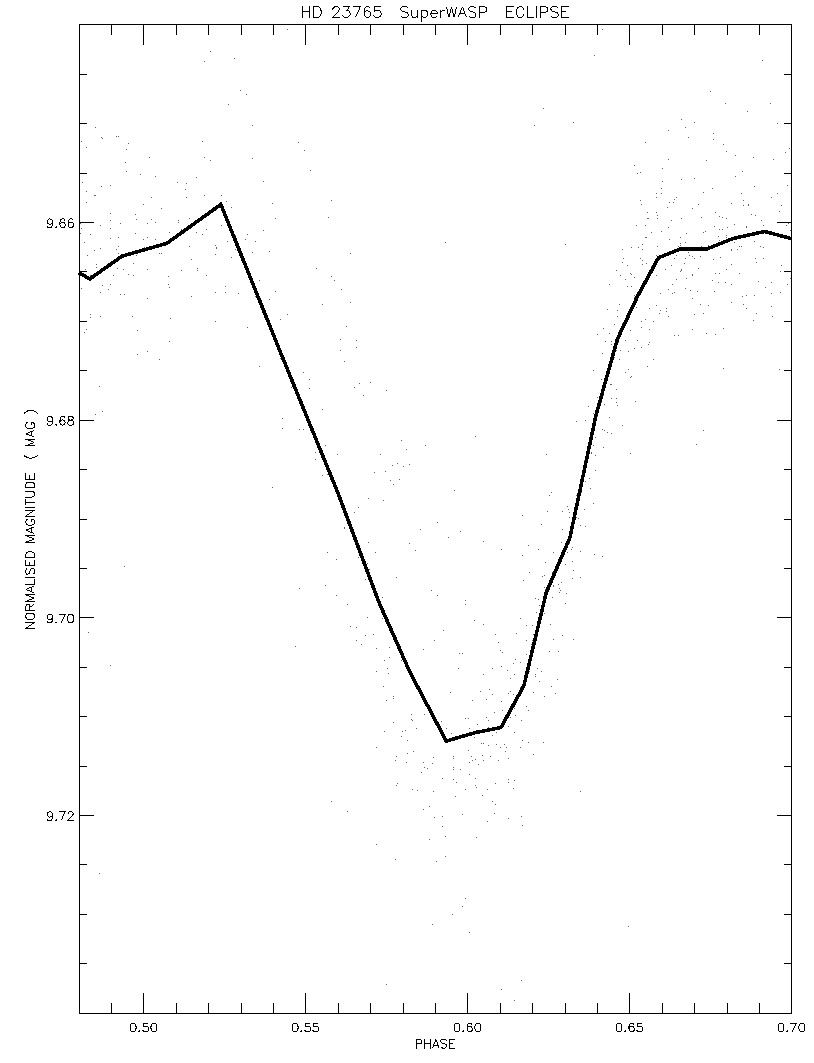}}
\resizebox{3cm}{!}{\includegraphics{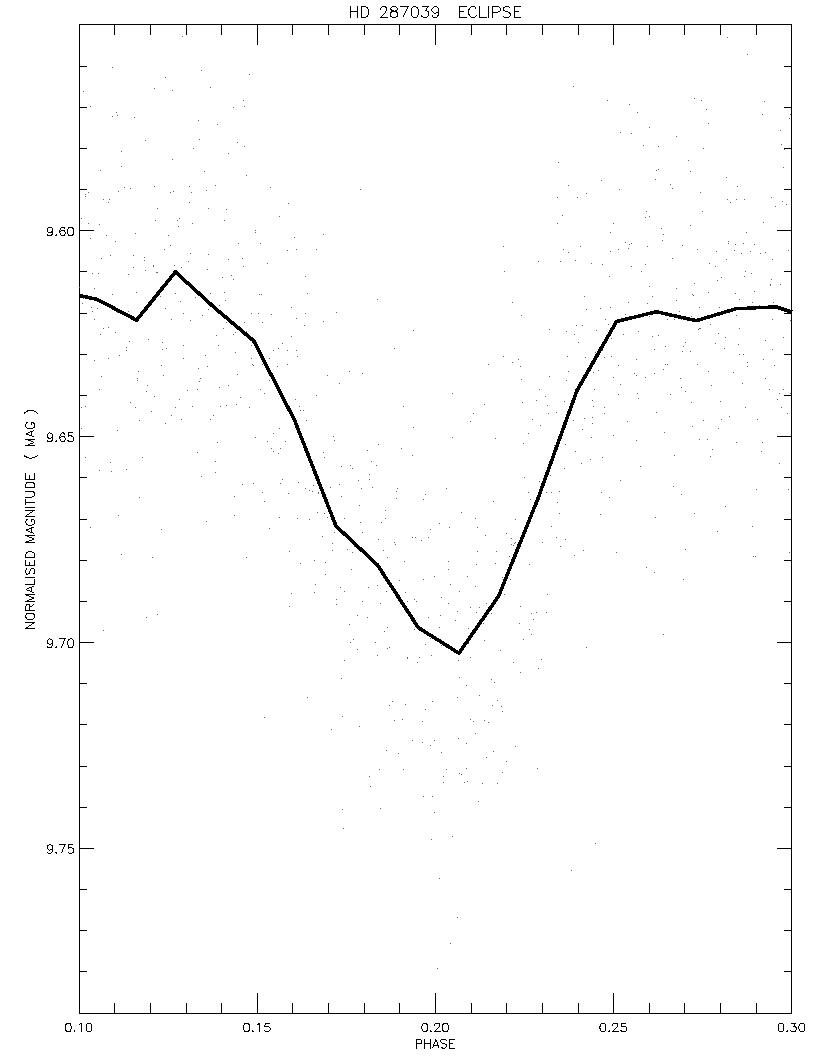}}
\resizebox{3cm}{!}{\includegraphics{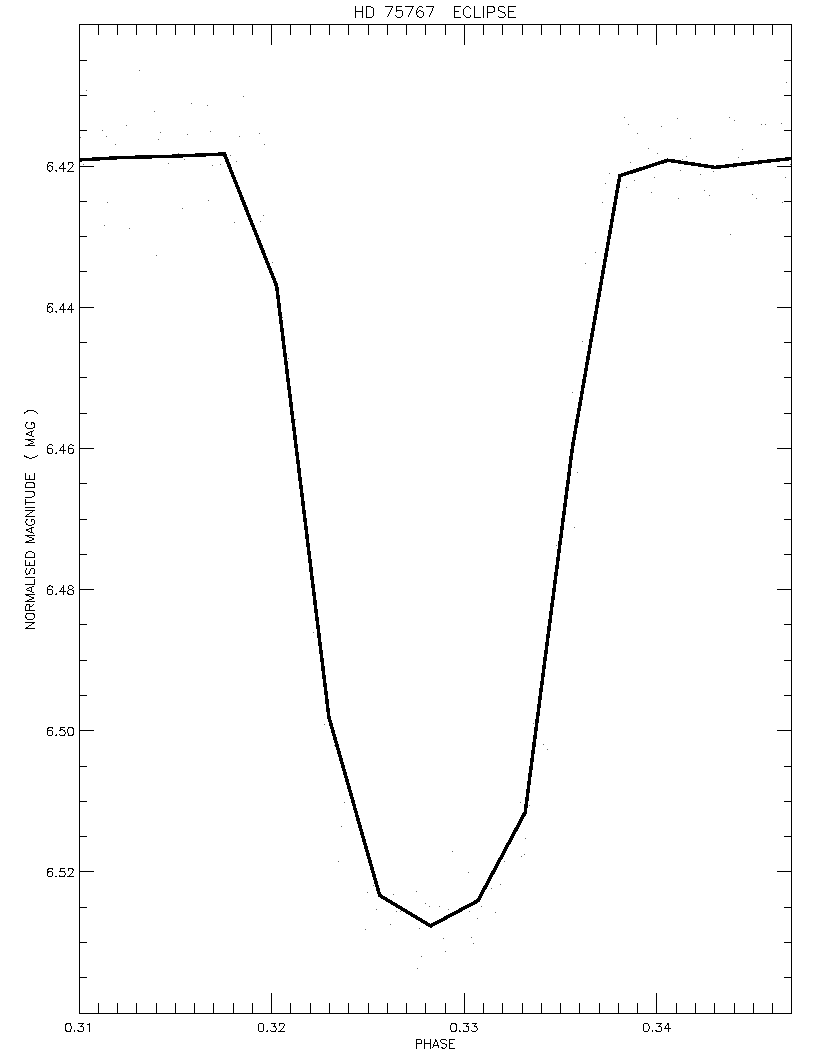}}
\resizebox{3cm}{!}{\includegraphics{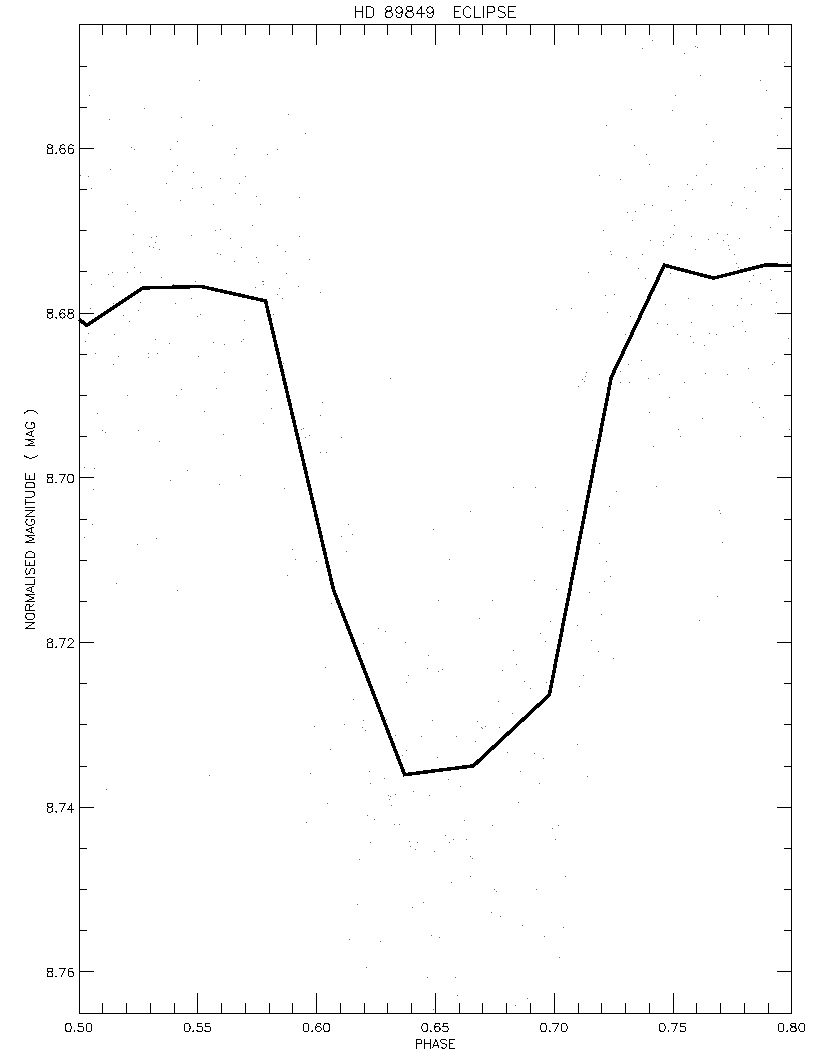}}
\caption{Close-up view of the eclipses of the low mass EB candidates.  Top row: BD-07~3648, HD~198044, HD~205403, HD~213597 and HD~222891.  Bottom row: HD~23765 (\textit{STEREO}/HI-1), HD~23765 (SuperWASP), HD~287039, HD~75767 and HD~89849.  The best-fitting lines shown bin 40 data points together with outliers more than 3~$\sigma$ from the mean of each bin culled, except for HD~198044 and HD~205403 (both 20 points), HD~75767 (10 points), HD~222891 (30 points) and HD~287039 (50 points).}
\label{fig5}
\end{figure*}

\noindent
Stars are listed in Table \ref{tab:1} in order of increasing radius of the companion.  Errors on the eclipse depths have been determined as the standard deviation of the out-of-eclipse lightcurves.  The period analysis was done in  \textsc{Peranso}\footnote{http://www.peranso.com}.  The Modified Julian Date (MJD-2400000) of mid-eclipse is given for each candidate.  Figure \ref{fig4} shows the lightcurves of the candidates phase-folded on the periods given and Figure \ref{fig5} shows close-ups of the eclipses, to help clarify which are most likely to be total, since for grazing eclipses the radii found will be lower limits.  Table \ref{table2} gives the times when the stars in Table \ref{tab:1} will next be viewed by \textit{STEREO}/HI-1, to facilitate observers who may wish to make observations simultaneously.

\subsection{Determination of radii}
\label{sec:findrad}

\noindent
One of the most crucial pieces of information which can be precisely extracted from an EB system is the radius of the companion star, given the 
knowledge of the eclipse depth and host star radius. The radii 
of the stars in our sample were estimated from the stellar temperature and luminosity, using the Stefan-Boltzmann relation, shown in Equation \ref{eq:stefb}, where $R$ is radius, $L$ is luminosity and $T_{eff}$ is effective temperature.  It was assumed in all cases that the observed eclipses are total (thus the radii derived are lower limits) and the radius of the companion then follows geometrically from the radius of the primary.
\begin{equation}
\frac{R_{\ast}}{R_{\odot}} = \sqrt{\frac{L_{\ast}}{L_{\odot}} \times \left(\frac{T_{eff \odot}}{T_{eff \ast}}\right)^{4}}
\label{eq:stefb}
\end{equation}

Only one of the stars in our sample has previously been analysed spectroscopically, HD~75767,
therefore it was decided to determine $\ensuremath{T_{\mathrm{eff}}}$
photometrically.  HD~75767 has been estimated to have a $\ensuremath{T_{\mathrm{eff}}}$ of 5823~K and a bolometric absolute magnitude of 4.60~mag in \citet{mishenina2009}. For all of our stars Johnson (\textit{B} and \textit{V} from 
\textsc{Simbad}), 2MASS 
\citep[\textit{J}, \textit{H} and $K_{s}$ from ][]{skrutskie2006} and Tycho 
\citep[$B_{T}$ and $V_{T}$ from][]{hog2000} photometry were available.
The photometric calibrations by \citet{casagrande2010} were then used
to estimate $\ensuremath{T_{\mathrm{eff}}}$ from various colour indices. Where 
available Str\"omgren photometry from \citet{hauck1998} was also used with 
calibrations by \citet{casagrande2010}, \citet{moon1985}, 
\citet{napiwotzki1993}, \citet{balona}, \citet{ribas}, and \citet{castelli}.
The effective temperature finally adopted for each star, listed in 
Table \ref{tab:1}, is the median value obtained from each colour 
whilst the uncertainty given is the standard deviation from the mean.

To derive a precise value of the stellar luminosity it is crucial to have a 
precise measurement of the stellar distance.  We have restricted our luminosity estimates to only those available in the new reduction of 
the HIPPARCOS catalog, presented by \citet{vanleeuwen2007}. For these stars 
the stellar luminosity was derived from the \textit{V} Johnson magnitude using 
the bolometric correction by \citet{balona}, the effective temperature
previously obtained, and reddening by \citet{amores2005}. The uncertainty on the 
luminosity was derived taking into account the uncertainties on the distance given by \citet{vanleeuwen2007}, an uncertainty in the bolometric correction of about 0.07 mag, and a reddening uncertainty of 0.01 mag \citep{fossati2008}.

For those stars not present in the new reduction of the HIPPARCOS catalog it was
decided just to give a rough estimate of the stellar radius on the basis of 
the effective temperature derived from the photometry \citep[stars marked with 
a $\ast$ in Table \ref{tab:1}. For this operation the values used are those given by][]{gray1992}. As the stellar radius was just a rough estimate, for these stars no uncertainty was derived, although an error bar of at least 50\% is plausible.  Therefore these values have to be taken with caution.

\onecolumn

\scriptsize
\begin{landscape}
\begin{table*}
\caption[]{ Comparison of low mass EB candidates observed by \textit{STEREO}/HI-1 in order of increasing radius of the companion.  The spectral type is as given in \textsc{Simbad} but where this has not been determined it is listed as NA.  The eclipse depth of each is given in mmag with an error corresponding to the standard deviation of the out-of-transit lightcurve, as found from the phase-folded lightcurves shown in Figure \ref{fig4}.  The Modified Julian Date (MJD-2400000) of the mid-eclipse time is given from the period analysis, which also provides the periods and their errors.  The $\ensuremath{T_{\mathrm{eff}}}$ and its error for each star was determined photometrically as described in Section \ref{sec:findrad}, with the median value given here and the error being the standard deviation.  Radii marked with a $\ast$ are estimated on the basis of $\ensuremath{T_{\mathrm{eff}}}$ rather than calculated from photometry as shown in Section \ref{sec:findrad}.  Note that the companion radii are lower limits and assume that the eclipses are total.  In the last column, each candidate is flagged as to whether it is previously known to be eclipsing from the analysis of \citet{wraight2011}.}
\protect\label{tab:1}
\begin{tabular}{llllllllllll}
\hline
Identity & RA & DEC & Mag & Sp type & Depth & MJD & Period & $\ensuremath{T_{\mathrm{eff}}}$ & $\mbox{R}_{\mbox{host}}$ & $\mbox{R}_{\mbox{comp}}$ & Known \\
 & (deg) & (deg) & (\textit{V}) &  & (mmag) &  & (days) & (K) & ($R_{\odot}$) & ($R_{\odot}$) &  \\
\hline
HD23765 & 57.1205 & 21.7975 & 9.53 & F8 & $49 \pm 17$ & 54213.264019 & $1.6865 \pm 0.0004$ & $5746 \pm 142$ & $0.98\ast$ & $\geq 0.22$ & no \\
\hline
HD287039 & 71.5819 & 12.7448 & 9.88 & F8 & $75 \pm 32$ & 54224.548593 & $2.2111 \pm 0.0004$ & $5310 \pm 145$ & $0.84\ast$ & $\geq 0.23$ & no \\
\hline
HD89849 & 155.541 & 6.21829 & 9.15 & F8 & $54 \pm 23$ & 54305.226101 & $3.0781 \pm 0.0017$ & $5789 \pm 83$ & $1.0\ast$ & $\geq 0.23$ & yes \\
\hline
BD-07 3648 & 203.727 & -8.44247 & 11.3 & NA & $472 \pm 131$ & 54353.170536 & $2.5265 \pm 0.0004$ & $4074 \pm 918$ & $0.35 \pm 0.09$ & $\geq 0.24 \pm 0.36$ & no \\
\hline
HD75767 & 133.068 & 8.06293 & 6.59 & G0+M3 & $103 \pm 13$ & 54285.947481 & $10.2478 \pm 0.0043$ & $5649 \pm 245$ & $1.02 \pm 0.09$ & $\geq 0.33 \pm 0.12$ & no \\
\hline
HD198044 & 312.091 & -22.7407 & 7.2 & F7V & $53 \pm 12$ & 54450.509489 & $5.15195 \pm 0.0015$ & $6116 \pm 236$ & $1.46 \pm 0.11$ & $\geq 0.34 \pm 0.13$ & yes \\
\hline
HD205403 & 323.766 & -3.7349 & 8.02 & F5 & $57 \pm 14$ & 54123.634818 & $2.4449 \pm 0.0005$ & $6771 \pm 136$ & $1.46\ast$ & $\geq 0.35$ & yes \\
\hline
HD213597 & 338.136 & 1.58245 & 7.81 & F0 & $33 \pm 13$ & 54138.602903 & $2.4238 \pm 0.0007$ & $6925 \pm 112$ & $1.96 \pm 0.06$ & $\geq 0.36 \pm 0.12$ & yes \\
\hline
HD222891 & 356.162 & -8.84879 & 8.07 & F8 & $39 \pm 12$ & 54150.650434 & $1.59495 \pm 0.0004$ & $6409 \pm 112$ & $1.80 \pm 0.06$ & $\geq 0.36 \pm 0.11$ & yes \\
\hline
\end{tabular}
\end{table*}

\begin{table*}
\caption[]{\small{The dates shown (in MJD-2400000) in this table are when the stars will next be observed by \textit{STEREO}/HI-1, until the two \textit{STEREO} spacecraft have passed behind the Sun.  The times take into account the mask that is routinely applied to exclude data likely to be affected by solar activity.  As the Sun is going to be at maximum during these observations, it is not expected that observations will be reliable within the region of the CCDs excluded by the mask.  The designation NA indicates that data will not be available for that star on that occasion, in particular HD~205403 is currently just outside the field of view of the \textit{STEREO}/HI-1B imager and will not be visible during the coming orbits for that satellite.  Observation periods that may be occurring when the relevant spacecraft are near to losing communication due to being behind the Sun are marked with a $\ast$ and may not be reported, or suffer from a reduced cadence and missing data.}} 
\protect\label{table2} 
\begin{tabular}{|c|cccccc|}
\hline
Identity & \multicolumn{6}{|c|}{start and end dates of upcoming \textit{STEREO}/HI-1 observations (MJD-2400000)}\\ \cline{2-7}
  & \textit{STEREO}/HI-1A & \textit{STEREO}/HI-1B & \textit{STEREO}/HI-1A & \textit{STEREO}/HI-1B & \textit{STEREO}/HI-1A & \textit{STEREO}/HI-1B\\
\hline
HD~23765 & 56280.01953 - 56293.18359 & 56202.00651 - 56217.03385 & 56624.57552 - 56637.73958 & 56591.04297 - 56606.07031 & 56969.13151 - 56982.29557 & 56980.07943 - 56995.10677  \\
HD~287039 & 56291.18359 - 56310.37891 & 56239.84375 - 56259.09375 & 56635.68359 - 56654.87891 & 56642.99219 - 56662.24219 & 56980.18359 - 56999.37891 & 57046.14063 - 57065.39063$\ast$ \\
HD~75767 & 56349.59766 - 56368.34766 & 56279.20508 - 56302.45508 & 56694.10677 - 56712.85677 & 56668.17578 - 56691.42578 & 57038.61589 - 57057.36589 & NA \\
HD~89849 & 56370.43750 - 56382.88281 & 56312.36328 - 56327.25000 & 56714.96615 - 56727.41146 & 56701.26563 - 56716.15234 & 57059.49479 - 57071.94010 & 57090.16797 - 57105.05469$\ast$ \\
BD-07~3648 & 56417.92969 - 56431.03906 & 56368.05469 - 56382.02734 & 56762.46615 - 56775.57552 & 56756.98438 - 56770.95703 & NA & 57145.91406 - 57159.88672 \\
HD~198044 & 56172.34766 - 56185.67969 & 56473.01367 - 56486.29102 & 56516.87500 - 56530.20703 & 56861.95703 - 56875.23438 & 56861.40234 - 56874.73438 & 57250.90039 - 57264.17773$\ast$ \\
HD~205403 & 56187.16016 - 56204.74609 & NA & 56531.19922 - 56548.78516 & NA & 56875.23828 - 56892.82422 & NA \\
HD~213597 & 56202.41016 - 56221.57813 & 56496.00391 - 56514.67188 & 56546.53255 - 56565.70052 & 56884.10156 - 56902.76953 & 56890.65495 - 56909.82292 & NA \\
HD~222891 & 56214.27344 - 56225.60547 & 56124.37891 - 56141.98828 & 56558.38672 - 56569.71875 & 56513.87891 - 56531.48828 & 56902.50000 - 56913.83203 & 56903.37891 - 56920.98828 \\
\hline
\end{tabular}
\end{table*}
\end{landscape}
\normalsize
\twocolumn

\subsection{Comments on individual candidates}

\subsubsection{HD 23765}

\noindent
This star is well isolated and the archival SuperWASP observations confirm the eclipse depth and show that only one very faint star remains as a potential contaminant.  The host star is faint, for \textit{STEREO} to observe such low amplitude variability, and detail on the shape of the eclipse is poorly defined but the SuperWASP data shows clear u-shaped eclipses, along with evidence of slight tidal distortion in the host.  Given the spectral type of the host and the short period, there is the potential for this to be due to a very low mass companion.  The SuperWASP lightcurve is comparatively noisy and some cleaning was required in \textsc{Peranso}, with slight indications that this might be partly due to very long period variability on a timescale greater than one year, although this would require more detailed observations to confirm or refute.  Undetrended \textit{STEREO} observations do not show this but there are hints of sinusoidal variability with a period about seven times the orbital period, however \textit{STEREO}/HI-1B data could not be included in this check for longer period variability due to interference from several de-pointing events.

\subsubsection{HD 287039}

\noindent
An apparently isolated star, there appear to be no candidates that could contaminate the lightcurve.  The eclipses are fairly deep, being 75~mmag, compared to others in the sample but also of short duration and the ingress and egress phases appear sharp.  The $\ensuremath{T_{\mathrm{eff}}}$ of 5310~K estimated here results in a lower estimated radius for the host than all but one of the other stars in the sample.  The eclipses are therefore consistent with a companion of 0.23~$R_{\odot}$.  There is no sign of tidal distortion in the host, despite the short period, and the companion may therefore be very low mass.

\subsubsection{HD 89849}

\noindent
This star is well isolated with only a single very faint star near enough to potentially contaminate the lightcurve.  The eclipses are box-like but with sufficient depth (54~mmag) and clear tidal distortion in the host that the companion is expected to be in the low mass stellar regime.  The radius estimated for the companion is 0.23~$R_{\odot}$, in agreement with this conclusion.  However, this is dependent upon the accuracy of the estimated 1~$R_{\odot}$ for the primary, which has no parallax but a $\ensuremath{T_{\mathrm{eff}}}$ of 5789~K.

\subsubsection{BD-07~3648}

\noindent
For this star, the photometry used in calculating the effective temperature was taken from the NOMAD1 catalogue \citep{zacharias2004}.  The star was initially suspected as being a low mass system due to the large parallax of $35.4 \pm 19.6~{ \mbox{mas}}$ \citep{vanaltena1995} combined with the faintness of the object.  Unlike the other stars in this sample, however, the large eclipse depth indicates that the secondary has a similar size to the primary and therefore that the photometry might include a significant component of light from this star also. From the $B-V$ colour we estimate an effective temperature of about 6000\,K.  Making use of the available 2MASS photometry we obtained three comparable values of $\ensuremath{T_{\mathrm{eff}}}$ around 4000\,K, therefore we finally adopted a $\ensuremath{T_{\mathrm{eff}}}$ of 4074$\pm$918\,K. From the transit depth, we know that the two stars have rather similar radii, meaning that the photometry might be affected by the secondary star, thereby introducing systematic errors in the adopted effective temperature. Taking into account the given parallax and the adopted $\ensuremath{T_{\mathrm{eff}}}$ we estimate the stellar radius to be $\sim$0.3\,$R_{\odot}$. However, contamination of the photometry by the light of the secondary star would result in an overestimation of the luminosity and temperature and, following from this, of the stellar radius, although the estimated radius does not seem unreasonable. Given the little information available and the complexity of the system, we are able just to give an indication of the spectral type of the two stars. Most likely the primary component is a late K-type star, while the secondary would then be an early M-type star. As a check, we attempted a fit of synthetic spectral energy distributions (SED), calculated with MARCS models \citep{gustafsson2008}, to the available photometry. Using a single SED the best fit is obtained for a dwarf star with a $\ensuremath{T_{\mathrm{eff}}}$ of about 4250\,K, which is then improved by adding a second SED for a dwarf star with a $\ensuremath{T_{\mathrm{eff}}}$ of about 3800\,K, in good agreement with our conclusions. As a further check, we attempted to fit the photometry assuming a much hotter primary star with a $\ensuremath{T_{\mathrm{eff}}}$ of 6000\,K, as suggested by the $B-V$ color. In this case, the fit of the infrared photometry was obtained only by assuming that the secondary star is a cool star with a radius of $\sim$12\,$R_{\odot}$, which is impossible given the observed transit depth, confirming therefore our conclusions. This would be a rather challenging system to confirm with spectroscopy or radial velocity studies. Nevertheless, as a nearby eclipsing binary of two very low mass stars it is more amenable to follow-up than the more distant systems of this type.  Secondary eclipses are just visible in the phase-folded lightcurve, showing no eccentricity. The secondary eclipse depth is hard to gauge, being visually of the order of 50~mmag deep whereas the scatter on the lightcurve is of the order of 0.35~mag.

\subsubsection{HD~75767}

\begin{figure*}
\centering
\resizebox{16cm}{!}{\includegraphics{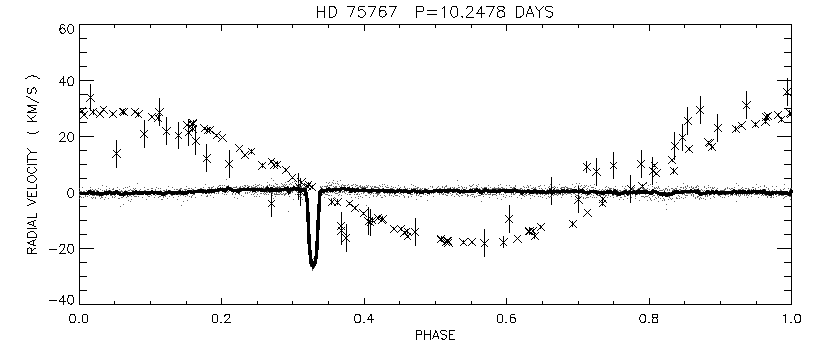}}
\caption{Radial velocity data of HD~75767 collected from \citet{griffin1991} and \citet{fuhrmann2005} with the phase-folded lightcurve of \textit{STEREO}/HI-1 overlaid in phase and set to 0 km/s.  The errors on the RV data points are the rms residuals given in \citet{griffin1991} and as specified for each data point from \citet{fuhrmann2005}.  The larger errors are from photograhic observations originating in \citet{sanford1931}.}
\label{fig6}
\end{figure*}

\noindent
This star is a known spectroscopic binary with an orbital period of $10.248042 \pm 0.000011$ days, which has been determined by radial velocity measurements \citep{griffin2006}.  The $\ensuremath{T_{\mathrm{eff}}}$ and absolute bolometric magnitude obtained from spectroscopic measurements in \citet{mishenina2009} are $5823 \pm 4$~K and 4.60, respectively, which are in agreement with our photometric determination of $5649 \pm 245$~K and 4.55.  The two stars are given to be spectral types G0 and M3 and this agrees very well with the radii found from this study.  The cause of a brightening in the phase-folded lightcurve around the primary eclipses is a systematic effect due to the polynomial detrending, as illustrated by Figure \ref{fig2}.  The radial velocity data clearly indicates that the eccentricity is very small at $0.0134 \pm 0.0020$ but secondary eclipses are not observed.  HD~75767 is actually a quadruple system, with the spectroscopic binary central pair orbited by a more distant M dwarf pair \citep{fuhrmann2005}.  This star has been known to be a spectroscopic binary for a long time \citep{sanford1931} but the eclipses have not been observed prior to \textit{STEREO}/HI-1, probably because of their shallow depth and short duration, the long period and also a lack of suitable reference stars for ground-based differential photometry.  Figure \ref{fig6} shows the radial velocity data collected by \citet{griffin1991} and \citet{fuhrmann2005} for this star with the \textit{STEREO}/HI-1 phase-folded lightcurve overlaid.  There appears from Figure \ref{fig6} to be a small discrepancy in the phase of the radial velocity measurements collected by \citet{griffin1991} and those of \citet{fuhrmann2005} and it is tempting to speculate that this is due to the influence of the circumbinary red dwarf binary reported by \citet{fuhrmann2005}.  The earliest radial velocity measurements date back to the 1920's \citep{sanford1931} and it is not inconceivable that the circumbinary stars might perturb the system in this way on decadal timescales.  If this is the case, then transit timing variations would confirm it as well as providing orbits and masses for both binaries.  It would also be worth monitoring for changes in eclipse depth in case the inclination of the system is similarly changing due to the combined motion of the two binaries.

\subsubsection{HD~198044}

\noindent
This star has been reported by \citet{samus2012} as a suspected variable with an amplitude of 50mmag, although the type of variability is not specified.  It appears to be very isolated, with no obvious candidates to contaminate the lightcurve.  Archival SuperWASP observations of this star have been made but are sparsely sampled in phase and the eclipses are not observed.  This is the second-brightest star in the sample and also has the second-longest period, factors which make observing with SuperWASP difficult.  The ingress and egress phases of the eclipses are very short indeed but there is some tidal distortion in the host, placing the companion in the low mass stellar regime.  This star is one of the closest in the sample for those with a parallax available, at just $50 \pm 1$ parsecs.

\subsubsection{HD 205403}

\noindent
This star is quite well isolated with only a single star almost two and a half arc-minutes away potentially contaminating the lightcurve, however this is almost exactly three magnitudes fainter in $R$.  The eclipses appear to be total and the ingress and egress phases are very short but some tidal distortion is evident in the host, making the companion most likely to be low mass star.  It is unclear whether or not secondary eclipses are seen for this star, as an extremely faint feature in the lightcurve could be easily confused with the signature caused by tidal distortion.

\subsubsection{HD 213597}

\noindent
As detailed in \citet{wraight2011}, this well-isolated star shows eclipses due to a potential substellar mass companion.  There appears to be an indication of tidal distortion in the host but the very shallow, box-like transits suggest the companion to be very close to the brown dwarf regime.  Interestingly, this is an early-type star with a slow projected rotational velocity, $V \sin i = 40 \pm 4~{ \mbox{km s}}^{-1}$ \citep{glebocki2005}.  It is therefore a strong candidate to be a chemically peculiar star.  There are, currently, only two known transiting planetary companions of early-type stars, WASP-33b \citep{smith2011} and KOI-13 \citep{szabo2011}, both of which are much fainter than this.  A radius measurement has been made for this star, giving $r = 2.039 \pm 0.303 R_{\odot}$ (Masana, Jordi \& Ribas 2006).  There is a hint of asymmetry in the transit, although this needs to be confirmed with more sensitive observations.  Using the radius measurement above, we derive the radius of the companion to be $0.35 \pm 0.31 R_{\odot}$.  The host radius determined here, photometrically, is $1.96 \pm 0.06 R_{\odot}$, from which we derive the radius of the companion to be $0.36 \pm 0.12 R_{\odot}$, which agrees with this.  Given this apparent radius, a brown dwarf companion is unlikely, however this object is in close orbit of a star with $\ensuremath{T_{\mathrm{eff}}}$ of nearly 7000~K and could easily have become inflated as a result.  Radial velocity measurements from \citet{nordstroem1997} show a reflex motion of just over $60~{ \mbox{km s}}^{-1}$, which suggests a low mass companion above the brown dwarf mass range.

\subsubsection{HD 222891}

\indent
A very isolated star, there is only one potential contaminant and this is almost four magnitudes fainter, so there is very good confidence in both the identity of the eclipsing variable and of the observed eclipse depth of 39~mmag.  The eclipses are of short duration but, although the shape is not clear, both the matched filter model and a close-up of the eclipse (Figure \ref{fig3}, right plot, and Figure \ref{fig5}, top-right plot, respectively) suggest the eclipses are most likely total.  More sensitive observations are required to confirm this and to search for secondary eclipses, which are not observed.  There is a faint hint of tidal distortion in the host but, given its spectral type of F8 and the observed eclipse depth, the companion is expected to be very low mass.

\section{\uppercase{Conclusions}}
\label{sec:discussion}

\noindent
The lack of bright eclipsing low mass, brown dwarf and exoplanet host stars is a major obstacle to detailed analysis of these interesting objects and to advancing the understanding of stellar formation and evolution.  \textit{STEREO} is an ideal platform for carrying out a survey for such objects and whilst the sample of nine presented here consists of companions likely to be too massive to be exoplanets, all are likely to be stars of very late spectral type (M or late K).  In particular, BD-07~3648 appears to be one of the brightest detached eclipsing binaries where both stars are in the low mass regime.  If follow-up observations show any of these candidates to have masses in the brown dwarf regime, however, it would be a big step forward for research into brown dwarfs and their atmospheres and other characteristics.  Follow-up observations are required to ascertain the nature of all the candidates presented here, except for HD~75767 for which a wealth of data is available (Figure \ref{fig6}).  There are relatively few detached low mass EB systems, with just six previously known double-lined detached eclipsing binaries where the companion has a radius less than 0.4~$R_{\odot}$ \citep{feiden2012}.  This sample of nine bright low mass EB candidates represents a considerable improvement in both the statistics and variety of low mass stars which can be summed up as follows:
\begin{itemize}
\item With a brightness range of $V=6.59$~mag to $V=11.3$~mag, they are a sample suitable to be studied in great detail, refining parameters of both components with unprecedented precision.
\item A range of periods have been observed from $1.59495$~days to $10.2478$~days, that shows low mass stars can be found in a diverse range of environments and provides a new benchmark against which to test models of stellar formation and magnetic field generation.
\item The host stars for these EBs have a diverse range of temperatures, from 6925~K to 4074~K, providing an excellent comparison for models of stellar evolution.
\item Spectral types for the host stars range from F0 to late K, including an early-type star (HD~213597) and a star that itself appears to be low mass (BD-07~3648), reinforcing the observation that low mass stars can be found in almost any environment.
\item The lower limits of the radii of the companions (and also the primary for BD-07~3648) are determined to be in the range 0.22~$R_{\odot}$ to 0.36~$R_{\odot}$, providing 10 stars in bright eclipsing systems that are expected to be near to the boundary where stars become fully convective, challenging models of stellar structure to fully explain their properties.
\end{itemize}

\section*{\uppercase{Acknowledgements}}
\label{sec:acks}

\noindent The Heliospheric Imager (HI) instrument was developed by a collaboration that included the Rutherford Appleton Laboratory and the University of Birmingham, both in the United Kingdom, and the Centre Spatial de Li\'ege (CSL), Belgium, and the US Naval Research Laboratory (NRL), Washington DC, USA.  The \textit{STEREO}/SECCHI project is an international consortium of the Naval Research Laboratory (USA), Lockheed Martin Solar and Astrophysics Lab (USA), NASA Goddard Space Flight Center (USA), Rutherford Appleton Laboratory (UK), University of Birmingham (UK), Max-Planck-Institut f\"{u}r Sonnensystemforschung (Germany), Centre Spatial de Li\'ege (Belgium), Institut d'Optique Th\'eorique et Appliqu\'ee (France) and Institut d'Astrophysique Spatiale (France).  This research has made use of the \textsc{Simbad} database, operated at CDS, Strasbourg, France.  This research has made use of NASA's Astrophysics Data System.  This publication makes use of data products from the Two Micron All Sky Survey, which is a joint project of the University of Massachusetts and the Infrared Processing and Analysis Center/California Institute of Technology, funded by the National Aeronautics and Space Administration and the National Science Foundation.  This research has made use of version 2.31 \textsc{Peranso} light curve and period analysis software, maintained at CBA, Belgium Observatory http://www.cbabelgium.com.  This research has used data from the WASP public archive. The WASP consortium comprises of the University of Cambridge, Keele University, University of Leicester, The Open University, The Queen's University Belfast, St. Andrews University and the Isaac Newton Group. Funding for WASP comes from the consortium universities and from the UK's Science and Technology Facilities Council. KTW acknowledges support from a STFC studentship. LF is supported by a STFC rolling grant.  Many thanks to the anonymous referee for comments that have improved the quality of the paper.

\end{document}